\documentclass[12pt]{article}
\usepackage{a4wide,amsmath,amssymb,array}

\catcode`\_=12
\let\_=_
\catcode`\!=8

\newcommand{\ie}{i.e.\ }
\newcommand{\eg}{e.g.\ }

\makeatletter
\def\reportno#1{\gdef\@reportno{#1}}
\def\@maketitle{%
  \hfill{\small\begin{tabular}[t]{r}%
    \@reportno
  \end{tabular}\par}%
  \vskip 2em%
  \begin{center}%
    \let \footnote \thanks
    {\large \@title \par}%
    \vskip 1.5em%
    {
      \lineskip .5em%
      \begin{tabular}[t]{c}%
        \@author  
      \end{tabular}\par}%
    \vskip 1em%
    {
     \@date}%
  \end{center}%
  \par
  \vskip 1.5em}
\makeatother

\begin{document}

\reportno{MPP--2004--102\\hep--ph/0408283}

\title{SUSY Les Houches Accord I/O made easy}

\author{T. Hahn \\
Max-Planck-Institut f\"ur Physik \\
F\"ohringer Ring 6, D--80805 Munich, Germany}

\date{August 25, 2004}

\maketitle

\begin{abstract}
A library for reading and writing data in the SUSY Les Houches Accord 
format is presented.  The implementation is in native Fortran 77.  The 
data are contained in a single array conveniently indexed by 
preprocessor statements.
\end{abstract}


\section{Introduction}

The SUSY Les Houches Accord (SLHA) has standardized and significantly
simplified the exchange of input and output parameters of SUSY models
between such disparate applications as spectrum calculators and event
generators.

While the SLHA specifications \cite{slha} include the precise formats
for Fortran I/O, it is nevertheless not entirely straightforward to read
or write a file in SLHA format.  The present library provides the user
with simple routines to read and write files in SLHA format, as well as
a few utility routines.  One thing the library does not do is modify the
numbers, which means there is no routine to compute, say, a particular
quantity at a new scale.

Sect.\ \ref{sect:data} describes the organization of the data
structures, Sect.\ \ref{sect:ref} gives the reference information for
the library routines, Sect.\ \ref{sect:examples} shows the usage in some
examples, Sect.\ \ref{sect:build} contains download and build
instructions, and Sect.\ \ref{sect:summary} summarizes.


\section{Data structures}
\label{sect:data}

The SLHA library is written in Fortran 77.  All routines operate on a
double-precision array, \texttt{slhadata}, which is about the simplest
conceivable data format for this purpose in Fortran.  For convenience of
use, this array is accessed via preprocessor statements, so the user
never needs to memorize any actual indices for the \texttt{slhadata}
array.  A file containing the preprocessor definitions must thus be
included.

The \texttt{slhadata} array consists of a `static' part containing the
information from SLHA \texttt{BLOCK} sections and a `dynamic' part
containing the information from SLHA \texttt{DECAY} sections.  The
static part is indexed by preprocessor variables defined in
\texttt{SLHA.h}, the dynamic part is accessed through the
\texttt{SLHAGetDecay}, \texttt{SLHANewDecay}, and \texttt{SLHAAddDecay}
functions and subroutines (see Sect.\ \ref{sect:ref}).

In addition, descriptive names for the PDG codes of the particles are 
declared in \texttt{PDG.h}.  These are needed \eg to access the decay 
information.

\subsection{SLHA blocks}

The explicit indexing of the \texttt{slhadata} need not (and should not)
be done by the user.  Rather, the members of the SLHA data structure are
accessed through preprocessor variables.  Tables \ref{tab:para1},
\ref{tab:para2}, \ref{tab:para3}, and \ref{tab:para4} list the
preprocessor variables defined in \texttt{SLHA.h} which follow closely
the definition of the Accord \cite{slha}.  Note that preprocessor
symbols are case sensitive.

As far as there is overlap, the names for the block members have been
chosen similar to the ones used in the MSSM model file of
\textit{FeynArts} \cite{famssm}.  The following index conventions are
employed in the Tables:
\begin{alignat*}{2}
t &= 1\dots 4	& \qquad & \text{(s)fermion type:} \\
&		& & \qquad\text{1 = (s)neutrinos,} \\
&		& & \qquad\text{2 = isospin-down (s)leptons,} \\
&		& & \qquad\text{3 = isospin-up (s)quarks,} \\
&		& & \qquad\text{4 = isospin-down (s)quarks} \\[1ex]
g &= 1\dots 3	& & \text{(s)fermion generation} \\[1ex]
s &= 1\dots 2	& & \text{number of sfermion mass-eigenstate,} \\
&		& & \qquad\text{in the absence of mixing 1 = L, 2 = R} \\[1ex]
c &= 1\dots 2	& & \text{number of chargino mass-eigenstate} \\[1ex]
n &= 1\dots 4	& & \text{number of neutralino mass-eigenstate}
\end{alignat*}

Matrices have a ``\texttt{Flat}'' array superimposed for convenience, in 
Fortran's standard column-major convention, \eg
\texttt{USf(1,1)} $\equiv$ \texttt{USfFlat(1)},
\texttt{USf(2,1)} $\equiv$ \texttt{USfFlat(2)}, 
\texttt{USf(1,2)} $\equiv$ \texttt{USfFlat(3)}, 
\texttt{USf(2,2)} $\equiv$ \texttt{USfFlat(4)}.
This makes it possible to \eg copy such a matrix with just a single
do-loop.

\begin{table}
\begin{center}
\begin{tt}
\begin{small}
\begin{tabular}{|>{\normalfont\scshape}l|l|ll|} \hline
\textnormal{Block name} &
\textnormal{Offset and length} &
\textnormal{Members} & \\ 
\hline\hline
modsel	& OffsetModSel	& ModSel_Model	& \\
	& LengthModSel	& ModSel_Content & \\
	&		& ModSel_GridPts & \\
	&		& ModSel_Qmax & \\
	&		& ModSel_PDG($i$) & $i = 1\dots 5$ \\
\hline\hline
sminputs & OffsetSMInputs & SMInputs_AlfaMZ & \\
	& LengthSMInputs & SMInputs_GF	& \\
	&		& SMInputs_AlfasMZ & \\
	&		& SMInputs_MZ	& \\
	&		& SMInputs_Mf($t$) & $t = 2\dots 4$ \\
	&		& SMInputs_Mtau & $\equiv$ SMInputs_Mf(2) \\
	&		& SMInputs_Mt	& $\equiv$ SMInputs_Mf(3) \\
	&		& SMInputs_Mb	& $\equiv$ SMInputs_Mf(4) \\
\hline\hline
minpar	& OffsetMinPar	& MinPar_Q	& \\
	& LengthMinPar	& MinPar_M0	& \\
	&		& MinPar_Lambda	& $\equiv$ MinPar_M0 \\
	&		& MinPar_M12	& \\
	&		& MinPar_Mmess	& $\equiv$ MinPar_M12 \\
	&		& MinPar_M32	& $\equiv$ MinPar_M12 \\
	&		& MinPar_TB	& \\
	&		& MinPar_signMUE & \\
	&		& MinPar_A	& \\
	&		& MinPar_N5	& $\equiv$ MinPar_A \\
	&		& MinPar_cgrav	& \\
\hline
\end{tabular}
\end{small}
\end{tt}
\end{center}
\caption{\label{tab:para1}Preprocessor variables defined in
\texttt{SLHA.h} to access the \texttt{slhadata} array.}
\end{table}

\begin{table}
\begin{center}
\begin{small}
\begin{tt}
\begin{tabular}{|>{\normalfont\scshape}l|l|ll|} \hline
\textnormal{Block name} &
\textnormal{Offset and length} &
\textnormal{Members} & \\
\hline\hline
extpar	& OffsetExtPar	& ExtPar_Q	& \\
	& LengthExtPar	& ExtPar_M1	& \\
	&		& ExtPar_M2	& \\
	&		& ExtPar_M3	& \\
	&		& ExtPar_Af($t$) & $t = 2\dots 4$ \\
	&		& ExtPar_Atau	& $\equiv$ ExtPar_Af(2) \\
	&		& ExtPar_At	& $\equiv$ ExtPar_Af(3) \\
	&		& ExtPar_Ab	& $\equiv$ ExtPar_Af(4) \\
	&		& ExtPar_MHu2	& \\
	&		& ExtPar_MHd2	& \\
	&		& ExtPar_MUE	& \\
	&		& ExtPar_MA02	& \\
	&		& ExtPar_TB	& \\
	&		& ExtPar_MSL($g$) & $g = 1\dots 3$ \\
	&		& ExtPar_MSE($g$) & $g = 1\dots 3$ \\
	&		& ExtPar_MSQ($g$) & $g = 1\dots 3$ \\
	&		& ExtPar_MSU($g$) & $g = 1\dots 3$ \\
	&		& ExtPar_MSD($g$) & $g = 1\dots 3$ \\
	&		& ExtPar_N5($g$) & $g = 1\dots 3$ \\
\hline\hline
mass	& OffsetMass	& Mass_Mf($t$,$g$) & $t = 1\dots 4,$ \\
	& LengthMass	&		& $g = 1\dots 3$ \\
	&		& Mass_MSf($s$,$t$,$g$) & $s = 1\dots 2,$ \\
	&		& 		& $t = 1\dots 4,$ \\
	&		& 		& $g = 1\dots 3$ \\
	&		& Mass_MZ	& \\
	&		& Mass_MW	& \\
	&		& Mass_Mh0	& \\
	&		& Mass_MHH	& \\
	&		& Mass_MA0	& \\
	&		& Mass_MHp	& \\
	&		& Mass_MNeu($n$) & $n = 1\dots 4$ \\
	&		& Mass_MCha($c$) & $c = 1\dots 2$ \\
	&		& Mass_MGl	& \\
	&		& Mass_MGrav	& \\
\hline
\end{tabular}
\end{tt}
\end{small}
\end{center}
\caption{\label{tab:para2}Preprocessor variables defined in
\texttt{SLHA.h} to access the \texttt{slhadata} array (cont'd).}
\end{table}

\begin{table}
\begin{center}
\begin{small}
\begin{tt}
\begin{tabular}{|>{\normalfont\scshape}l|l|ll|} \hline
\textnormal{Block name} &
\textnormal{Offset and length} &
\textnormal{Members} & \\
\hline\hline
nmix	& OffsetNMix	& NMix_ZNeu($n!1$,$n!2$) & $n!1,n!2 = 1\dots 4$ \\
	& LengthNMix	& NMix_ZNeuFlat($i$)	& $i = 1\dots 16$ \\
\hline\hline
umix	& OffsetUMix	& UMix_UCha($c!1$,$c!2$) & $c!1,c!2 = 1\dots 2$ \\
	& LengthUMix	& UMix_UChaFlat($i$)	& $i = 1\dots 4$ \\
\hline
vmix	& OffsetVMix	& VMix_VCha($c!1$,$c!2$) & $c!1,c!2 = 1\dots 2$ \\
	& LengthVMix	& VMix_VChaFlat($i$)	& $i = 1\dots 4$ \\
\hline\hline
	&		& SfMix_USf($s!1$,$s!2$,$t$) & $s!1,s!2 = 1\dots 2,$ \\
	&		&			& $t = 2\dots 4$ \\
	&		& SfMix_USfFlat($i$,$t$) & $i = 1\dots 4,$ \\
	&		&			& $t = 2\dots 4$ \\
\hline
staumix	& OffsetStauMix	& StauMix_USf($s!1$,$s!2$) & $\equiv$ SfMix_USf($s!1$,$s!2$,2) \\
	& LengthStauMix	& StauMix_USfFlat($i$)	& $\equiv$ SfMix_USfFlat($i$,2) \\
\hline
stopmix	& OffsetStopMix	& StopMix_USf($s!1$,$s!2$) & $\equiv$ SfMix_USf($s!1$,$s!2$,3) \\
	& LengthStopMix	& StopMix_USfFlat($i$)	& $\equiv$ SfMix_USfFlat($i$,3) \\
\hline
sbotmix	& OffsetSbotMix	& SbotMix_USf($s!1$,$s!2$) & $\equiv$ SfMix_USf($s!1$,$s!2$,4) \\
	& LengthSbotMix	& SbotMix_USfFlat($i$)	& $\equiv$ SfMix_USfFlat($i$,4) \\
\hline\hline
alpha	& OffsetAlpha	& Alpha_Alpha	& \\
	& LengthAlpha	& 		& \\
\hline\hline
hmix	& OffsetHMix	& HMix_Q	& \\
	& LengthHMix	& HMix_MUE	& \\
	&		& HMix_TB	& \\
	&		& HMix_VEV	& \\
	&		& HMix_MA02	& \\
\hline\hline
gauge	& OffsetGauge	& Gauge_Q	& \\
	& LengthGauge	& Gauge_g1	& \\
	&		& Gauge_g2	& \\
	&		& Gauge_g3	& \\
\hline\hline
msoft	& OffsetMSoft	& MSoft_Q	& \\
	& LengthMSoft	& MSoft_M1	& \\
	&		& MSoft_M2	& \\
	&		& MSoft_M3	& \\
	&		& MSoft_MHu2	& \\
	&		& MSoft_MHd2	& \\
	&		& MSoft_MSL($g$) & $g = 1\dots 3$ \\
	&		& MSoft_MSE($g$) & $g = 1\dots 3$ \\
	&		& MSoft_MSQ($g$) & $g = 1\dots 3$ \\
	&		& MSoft_MSU($g$) & $g = 1\dots 3$ \\
	&		& MSoft_MSD($g$) & $g = 1\dots 3$ \\
\hline
\end{tabular}
\end{tt}
\end{small}
\end{center}
\caption{\label{tab:para3}Preprocessor variables defined in
\texttt{SLHA.h} to access the \texttt{slhadata} array (cont'd).}
\end{table}

\begin{table}
\begin{center}
\begin{small}
\begin{tt}
\begin{tabular}{|>{\normalfont\scshape}l|l|ll|} \hline
\textnormal{Block name} &
\textnormal{Offset and length} &
\textnormal{Members} & \\
\hline\hline
	&		& Af_Q($t$)	& $t = 2\dots 4$ \\
	&		& Af_Af($t$)	& $t = 2\dots 4$ \\
\hline
ae	& OffsetAe	& Ae_Q		& $\equiv$ Af_Q(2) \\
	& LengthAe	& Ae_Atau	& $\equiv$ Af_Af(2) \\
\hline
au	& OffsetAu	& Au_Q		& $\equiv$ Af_Q(3) \\
	& LengthAu	& Au_At		& $\equiv$ Af_Af(3) \\
\hline
ad	& OffsetAd	& Ad_Q		& $\equiv$ Af_Q(4) \\
	& LengthAd	& Ad_Ab		& $\equiv$ Af_Af(4) \\
\hline\hline
	&		& Yf_Q($t$)	& $t = 2\dots 4$ \\
	&		& Yf_Af($t$)	& $t = 2\dots 4$ \\
\hline
ye	& OffsetYe	& Ye_Q		& $\equiv$ Yf_Q(2) \\
	& LengthYe	& Ye_Atau	& $\equiv$ Yf_Yf(2) \\
\hline
yu	& OffsetYu	& Yu_Q		& $\equiv$ Yf_Q(3) \\
	& LengthYu	& Yu_At		& $\equiv$ Yf_Yf(3) \\
\hline
yd	& OffsetYd	& Yd_Q		& $\equiv$ Yf_Q(4) \\
	& LengthYd	& Yd_Ab		& $\equiv$ Yf_Yf(4) \\
\hline
\end{tabular}
\end{tt}
\end{small}
\end{center}
\caption{\label{tab:para4}Preprocessor variables defined in
\texttt{SLHA.h} to access the \texttt{slhadata} array (cont'd).}
\end{table}

\subsection{PDG particle identifiers}
\label{sect:pdg}

\texttt{PDG.h} defines the human-readable versions of the PDG codes
listed in Table \ref{tab:pdg}.  These are needed \eg to access the decay
information.  At run time, the subroutine \texttt{SLHAPDGName} can be
used to translate a PDG code into a particle name (see Sect.\
\ref{sect:pdgname}).

\begin{table}
\begin{center}
\begin{small}
\begin{tt}
\begin{tabular}[t]{|l|ll|l|} \hline
\textnormal{fermions} &
\textnormal{sfermions} & \\
\hline\hline
PDG_nu_e	& PDG_snu_e1	& PDG_snu_e2 \\
PDG_electron	& PDG_selectron1 & PDG_selectron2 \\
PDG_up		& PDG_sup1	& PDG_sup2 \\
PDG_down	& PDG_sdown1	& PDG_sdown2 \\
\hline
PDG_nu_mu	& PDG_snu_mu1	& PDG_snu_mu2 \\
PDG_muon	& PDG_smuon1	& PDG_smuon2 \\
PDG_charm	& PDG_scharm1	& PDG_scharm2 \\
PDG_strange	& PDG_sstrange1	& PDG_sstrange2 \\
\hline
PDG_nu_tau	& PDG_snu_tau1	& PDG_snu_tau2 \\
PDG_tau		& PDG_stau1	& PDG_stau2 \\
PDG_top		& PDG_stop1	& PDG_stop2 \\
PDG_bottom	& PDG_sbottom1	& PDG_sbottom2 \\
\hline
\end{tabular}
\begin{tabular}[t]{|l|l|} \hline
\textnormal{bosons} &
\textnormal{gauginos} \\
\hline\hline
PDG_h0		& PDG_neutralino1 \\
PDG_HH		& PDG_neutralino2 \\
PDG_A0		& PDG_neutralino3 \\
PDG_Hp		& PDG_neutralino4 \\
PDG_photon	& PDG_chargino1 \\
PDG_Z		& PDG_chargino2 \\
PDG_W		& PDG_gluino \\
PDG_gluon	& PDG_gravitino \\
PDG_graviton	& \\
\hline
\end{tabular}
\end{tt}
\end{small}
\end{center}
\caption{\label{tab:pdg}The PDG codes defined in \texttt{PDG.h}.}
\end{table}


\section{Routines provided by the SLHA library}
\label{sect:ref}

\subsection{SLHAClear}

\begin{verbatim}
        subroutine SLHAClear(slhadata)
        double precision slhadata(nslhadata)
\end{verbatim}
This subroutine sets all data in the \texttt{slhadata} array given as
argument to the value \texttt{invalid} (defined in \texttt{SLHA.h}).  It
is important that this is done before using \texttt{slhadata}, or else
any kind of junk that happens to be in the memory occupied by
\texttt{slhadata} will later on be interpreted as valid data.

\subsection{SLHARead}

\begin{verbatim}
        subroutine SLHARead(error, slhadata, filename, abort)
        integer error, abort
        double precision slhadata(nslhadata)
        character*(*) filename
\end{verbatim}
This subroutine reads the data in SLHA format from \texttt{filename} 
into the \texttt{slhadata} array.
If the specified file cannot be opened, the function issues an error 
message and returns \texttt{error = 1}.
The \texttt{abort} flag governs what happens when superfluous text is 
read, \ie text that cannot be interpreted as SLHA data.  If 
\texttt{abort} is 0, a warning is printed and reading continues.  
Otherwise, reading stops at the offending line and \texttt{error = 2} 
is returned.

\subsection{SLHAWrite}

\begin{verbatim}
        subroutine SLHAWrite(error, slhadata,
     &    program, version, filename)
        integer error
        double precision slhadata(nslhadata)
        character*(*) program, version, filename
\end{verbatim}
This subroutine writes the data in \texttt{slhadata} to
\texttt{filename}.  The name and version of the program that generates
the output is given in \texttt{program} and \texttt{version}.

\subsection{SLHAGetDecay}
\label{sect:getdecay}

\begin{verbatim}
        double precision function SLHAGetDecay(slhadata, parent_id,
     &    nchildren, child1_id, child2_id, child3_id, child4_id)
        implicit none
        double precision slhadata(*)
        integer parent_id
        integer nchildren, child1_id, child2_id, child3_id, child4_id
\end{verbatim}
This function extracts the decay
$$
\texttt{parent_id}\quad\to
\quad\texttt{child1_id}
\quad\texttt{child2_id}
\quad\texttt{child3_id}
\quad\texttt{child4_id}
$$
from the \texttt{slhadata} array, or the value \texttt{invalid} (defined
in \texttt{SLHA.h}) if no such decay can be found.  The parent and child
particles are given by their PDG identifiers (see Sect.\
\ref{sect:pdg}).  The return value is the total decay width if 
\texttt{nchildren = 0}, otherwise the branching ratio of the specified 
channel.

Note that only the first \texttt{nchildren} of the \texttt{child$n$_id}
are actually accessed and Fortran allows to omit the remaining ones in
the invocation (a strict syntax checker might issue a warning, though). 
Thus, for instance,
\begin{verbatim}
   Zbb = SLHAGetDecay(slhadata, PDG_Z, 2, PDG_bottom, -PDG_bottom)
\end{verbatim}
is a perfectly legitimate way to extract the $Z\to b\bar b$ decay.

\subsection{SLHANewDecay}

\begin{verbatim}
        integer function SLHANewDecay(slhadata, width, parent_id)
        double precision slhadata(nslhadata), width
        integer parent_id
\end{verbatim}
This function initiates the setting of decay information for the
particle specified by the \texttt{parent_id} PDG code, whose total decay
width is given by \texttt{width}.  The integer index it returns is
needed to subsequently add individual decay modes with
\texttt{SLHAAddDecay}.  If the fixed-length array \texttt{slhadata} 
becomes full, a warning is printed and zero is returned.  If a decay of 
the given particle is already present in \texttt{slhadata}, it is first 
removed.

\subsection{SLHAAddDecay}

\begin{verbatim}
        subroutine SLHAAddDecay(slhadata, br, decay,
     &    nchildren, child1_id, child2_id, child3_id, child4_id)
        double precision slhadata(nslhadata), br
        integer decay
        integer nchildren, child1_id, child2_id, child3_id, child4_id
\end{verbatim}
This subroutine adds the decay mode
$$
\texttt{(parent_id)}\quad\to
\quad\texttt{child1_id}
\quad\texttt{child2_id}
\quad\texttt{child3_id}
\quad\texttt{child4_id}
$$
to the decay section previously initiated by \texttt{SLHANewDecay}.
\texttt{decay} is the index obtained from \texttt{SLHANewDecay} (which
also sets the \texttt{parent_id}) and \texttt{child$n$_id} are the PDG
codes of the final-state particles.  The branching ratio is given in
\texttt{br}.  If the fixed-length array \texttt{slhadata} becomes full,
a warning is printed and \texttt{decay} is set to zero.

If \texttt{decay} is zero, an overflow of \texttt{slhadata} in an
earlier invocation is silently assumed and no action is performed.  It
is therefore sufficient to check for overflow only once, after setting
all decay modes (unless, of course, one needs to pinpoint the exact
location of the overflow).

As with \texttt{SLHAGetDecay} (see Sect.\ \ref{sect:getdecay}), only the
first \texttt{nchildren} of the \texttt{child$n$_id} are actually
accessed and Fortran allows to omit the remaining ones in the
invocation.

\subsection{SLHAExist}

\begin{verbatim}
        logical function SLHAExist(slhablock, length)
        double precision slhablock(*)
        integer length
\end{verbatim}
This function tests whether a given SLHA block is not entirely empty, 
\ie it returns \texttt{.TRUE.} if at least one member of the block is 
valid.  The SLHA blocks are most conveniently accessed using the 
\texttt{Offset...} and \texttt{Length...} definitions (see Sect.\ 
\ref{sect:data}), \eg
\begin{verbatim}
        if( SLHAExist(slhadata(OffsetMass), LengthMass) ) ...
\end{verbatim}

\subsection{SLHAValid}

\begin{verbatim}
        logical function SLHAValid(slhablock, length)
        double precision slhablock(*)
        integer length
\end{verbatim}
This function tests whether a given SLHA block consists entirely of
valid data, \ie it returns \texttt{.FALSE.} if at least one member of
the block is invalid.  The SLHA blocks are most conveniently accessed
using the \texttt{Offset...}  and \texttt{Length...} definitions (see
Sect.\ \ref{sect:data}), \eg
\begin{verbatim}
        if( SLHAValid(slhadata(OffsetNMix), LengthNMix) ) ...
\end{verbatim}

\subsection{SLHAPDGName}
\label{sect:pdgname}

\begin{verbatim}
        subroutine SLHAPDGName(code, name)
        integer code
        character*(PDGLen) name
\end{verbatim}
This subroutine translates a PDG code into a particle name.  The sign of 
the PDG code is ignored, hence the same name is returned for a particle 
and its antiparticle.  The maximum length of the name, \texttt{PDGLen}, 
is defined in \texttt{PDG.h}.


\section{Examples}
\label{sect:examples}

Consider the following example program, which just copies one SLHA file 
to another:
\begin{verbatim}
        program copy_slha_file
        implicit none

#include "SLHA.h"

        integer error
        double precision slhadata(nslhadata)

        call SLHAClear(slhadata)

        call SLHARead(error, slhadata, "infile.slha", 0)
        if( error .ne. 0 ) stop "Read error"

        call SLHAWrite(error, slhadata,
     &    "My Test Program", "1.0", "outfile.slha")
        if( error .ne. 0 ) stop "Write error"
        end
\end{verbatim}
Already in this simple program a couple of things can be seen:
\begin{itemize}
\item the file \texttt{SLHA.h} must be included in every function or 
      subroutine that uses the SLHA routines and this must be done
      using the preprocessor \texttt{\#include} (not Fortran's
      \texttt{include}), thus the program file should have the
      extension \texttt{.F} (capital F).
\item \texttt{slhadata} must be declared as a double-precision array of 
      length \texttt{nslhadata}.
\item One should not continue with processing if a non-zero error
      flag is returned.
\end{itemize}
A more sensible application would add something to the \texttt{slhadata} 
before writing them out again.  The next little program pretends to 
compute the fermionic Z decays (by calling a hypothetical subroutine 
\texttt{MyCalculation}) and adds them to \texttt{slhadata}:
\begin{verbatim}
        program compute_decays
        implicit none

#include "SLHA.h"
#include "PDG.h"

        integer error, decay, t, g
        double precision slhadata(nslhadata)
        double precision total_width, br(4,3)
        integer ferm_id(4,3)
        data ferm_id /
     &    PDG_nu_e, PDG_electron, PDG_up, PDG_down,
     &    PDG_nu_mu, PDG_muon, PDG_charm, PDG_strange, 
     &    PDG_nu_tau, PDG_tau, PDG_top, PDG_bottom /

        call SLHAClear(slhadata)

        call SLHARead(error, slhadata, "infile.slha", 0)
        if( error .ne. 0 ) stop "Read error"

* compute the decays with parameters taken from the slhadata:
        call MyCalculation(SMInputs_MZ, MinPar_TB, ...,
     &    total_width, br)

        decay = SLHANewDecay(slhadata, total_width, PDG_Z)
        do t = 1, 4
          do g = 1, 3
            call SLHAAddDecay(slhadata, br(t,g), decay,
     &        2, ferm_id(t,g), -ferm_id(t,g))
          enddo
        enddo

        call SLHAWrite(error, slhadata,
     &    "My Test Program", "2.0", "outfile.slha")
        if( error .ne. 0 ) stop "Write error"
        end
\end{verbatim}
Demonstrated here is the access of SLHA data (\texttt{SMInputs_MZ}, 
\texttt{MinPar_TB}) and the setting of decay information.


\section{Building and Compiling}
\label{sect:build}

The SLHA library package can be downloaded as a gzipped tar archive from 
the Web site \texttt{http://www.feynarts.de/slha}.  After unpacking the 
archive, change into the directory \texttt{SLHALib-1.0} and type
\begin{verbatim}
  ./configure
  make
\end{verbatim}
A simple demonstration program (\texttt{demo}, source code in
\texttt{demo.F}) is built together with the library \texttt{libSLHA.a}.

Compiling a program that uses the SLHA library is in principle equally
straightforward.  The only tricky thing is that one has to relax
Fortran's 72-column limit.  This is because even lines perfectly within
the 72-column range may become longer after the preprocessor's 
substitutions.  While essentially every Fortran compiler offers such an 
option, the name is quite different.  A glance at the man page should 
suffice to find out.  Here are a few common choices:
\begin{center}
\begin{tabular}{l|l|l}
Compiler & Platform/OS & Option name \\ \hline
g77 & any & \texttt{-ffixed-line-length-none} \\
pgf77 & Linux x86 & \texttt{-Mextend} \\
f77 & Tru64 Alpha & \texttt{-extend_source} \\
f77 & SunOS, Solaris & \texttt{-e} \\
fort77 & HP-UX & \texttt{+es}
\end{tabular}
\end{center}
To compile and link your program, add this option and
\texttt{-I}\textit{path} \texttt{-L}\textit{path} \texttt{-lSLHA} to the
compiler command line, where \textit{path} is the location of the SLHA
library, \eg
\begin{verbatim}
pgf77 -Mextend -I$HOME/SLHALib-1.0 myprogram.F -L$HOME/SLHALib-1.0 -lSLHA
\end{verbatim}

All externally visible symbols of the SLHA library start with the prefix 
\texttt{SLHA} and should thus pretty much avoid symbol conflicts.


\section{Summary}
\label{sect:summary}

The SLHA library presented here provides simple functions to read and
write files in SLHA format.  Data are kept in a single double-precision
array and accessed through preprocessor variables.  The library is
written in native Fortran 77 and is easy to build.  The source code is
openly available at \texttt{http://www.feynarts.de/slha} and is
distributed under the GNU Library General Public License.

The author welcomes any kind of feedback, in particular bug and 
performance reports, at hahn@feynarts.de.


\newcommand{\volyearpage}[3]{\textbf{#1} (#2) #3}
\newcommand{\cpc}{\textsl{Comp.\ Phys.\ Commun.} \volyearpage}

\end{document}